\documentclass[aps,prl,reprint,superscriptaddress, footinbib, longbibliography]{revtex4-2}
\usepackage{natbib}
\usepackage{graphicx}
\usepackage{dcolumn}
\usepackage{bm}
\usepackage{amsmath,amssymb, amsfonts}
\usepackage{subfigure}
\usepackage{ulem}
\usepackage{color}
\usepackage{xcolor}
\usepackage{float}
\usepackage{makecell}
\usepackage{newtxtext,newtxmath}
\usepackage{mathrsfs}
\usepackage{gensymb}
\usepackage{multirow}
\usepackage{booktabs}
\usepackage{threeparttable}
\usepackage{ulem}
\usepackage{rotating}

\usepackage{tabularx}
\usepackage{aas_macros}
\usepackage{verbatim}
\usepackage{url }
\usepackage{caption}
\usepackage{subcaption}
\graphicspath{{figure/}}
\usepackage[breaklinks,colorlinks,citecolor=blue]{hyperref}
\usepackage[mathlines]{lineno}

\newcommand{\dif}{{\rm{d}}}

\DeclareUnicodeCharacter{2212}{-}

\begin{document}

\preprint{APS/123-QED}


\title{Population statistics of  nanohertz gravitational wave sources}

\author{Jiming Yu}
\email{yjm8012@mail.ustc.edu.cn}
\affiliation{School of Physics and Optoelectronic Engineering, Yangtze University, Jingzhou, Hubei 434023, China}
\affiliation{Department of Astronomy, School of Physics and Astronomy, Shanghai Jiao Tong University, Shanghai, 200240, China}
\affiliation{Key Laboratory for Particle Astrophysics and Cosmology (MOE) / Shanghai Key Laboratory for Particle Physics and Cosmology, China}
\author{Zhen Pan}
\email{zhpan@sjtu.edu.cn}
\affiliation{Tsung-Dao Lee Institute, Shanghai Jiao-Tong University, Shanghai, 520 Shengrong Road, 201210, China}
\affiliation{Department of Astronomy, School of Physics and Astronomy, Shanghai Jiao Tong University, Shanghai, 200240, China}
\author{Xiao Xue}
\email{xxue@ifae.es}
\affiliation{Institut de F\'{i}sica d’Altes Energies (IFAE), The Barcelona Institute of Science and Technology,
Campus UAB, 08193 Bellaterra (Barcelona), Spain}
\author{Liang Dai}
\email{liangdai@fudan.edu.cn}
\affiliation{Center for Astronomy \& Astrophysics and Department of Physics, Fudan University, Shanghai 200438, P.R.China}

\date{\today}

\begin{abstract}
\noindent
 The recent detection of a nanohertz gravitational wave (GW) background by pulsar timing arrays (PTA) has sparked extensive discussions regarding its origin-whether it arises from astrophysical supermassive black hole binaries (SMBHBs) or from primordial GWs generated by various early universe processes. Previous studies suggest that a key discriminant between these two origins is the non-Gaussianity of the GW background prior to the detection of any individual source. In this Letter, we introduce a hierarchical Bayesian inference framework for inferring population properties of GW sources. This approach enables not only the measurement of evidence for different GW origins using PTA data but also the inference of population properties of astrophysical SMBHBs, by optimally leveraging non-Gaussian information in individual bright sources and in  power spectrum fluctuations of the GW background.
\end{abstract}

\maketitle

\noindent{\textit{\textbf{Introduction.}}}
Pulsar timing arrays (PTAs) probe gravitational waves (GWs) in the nanohertz (nHz) band by searching for correlated timing residuals in precisely monitored millisecond pulsars \cite{1975GReGr...6..439E, 1978SvA....22...36S, 1979ApJ...234.1100D, 1990ApJ...361..300F}. Recent PTA observations  \cite{2023ApJ...951L...8A, 2023A&A...678A..50E, 2023ApJ...951L...6R, 2023RAA....23g5024X, 2023MNRAS.519.3976M, 2025MNRAS.536.1467M} have reported evidence for a common-spectrum process with spatial correlations consistent with the Hellings \& Downs signature \cite{1983ApJ...265L..39H} of a stochastic gravitational wave background (GWB). These observations open a new low-frequency window on GW astronomy, but the physical origin of the signal remains uncertain. A population of supermassive black-hole binaries (SMBHBs) is a natural astrophysical explanation (see e.g. \cite{1995ApJ...446..543R, 2004ApJ...615...19E, 2004ApJ...611..623S, 2017MNRAS.464.3131K}), and predicts the characteristic strain $h_{\rm c}(f)\propto f^{-2/3}$ in the limit of circular, GW-driven binaries forming an effectively infinite population \cite{1963PhRv..131..435P, 1964PhRv..136.1224P, Maggiore:2007ulw}. Primordial processes in the early Universe, however, can also generate stochastic backgrounds in the PTA band \cite{2023ApJ...951L..11A}. Since astrophysical and primordial sources may produce GWBs of similar spectra over the currently accessible frequency range, identifying the origin of the signal requires information beyond the mean power spectrum density (PSD).

One such diagnostic is the non-Gaussianity (and anisotropy) of the background \cite{Mingarelli:2013dsa,NANOGrav:2023hvm,Wu:2024xkp, Raidal:2024tui, Ellis:2023dgf, Ellis:2023owy,2025PhRvD.111b3043S,2025PhRvD.111d3022X,2024ApJ...971L..10L,Falxa:2024ski,Ali-Haimoud:2026sbk, Hisamatsu:2026qvd,Domcke:2025esw,Konstandin:2024fyo,Bernardo:2024uiq}. An SMBHB background is generated by a finite number of discrete binaries rather than by a perfectly smooth random field. When many weak sources contribute to each frequency bin, their ensemble approaches a Gaussian background \cite{1997rggr.conf..373A, 1999PhRvD..59j2001A, 2017LRR....20....2R}. When a small number of bright binaries dominate, the signal develops non-Gaussian fluctuations and localized departures from a smooth power-law spectrum \cite{2016ApJ...819..163R, 2023PhRvD.107d3018A, Ellis:2023owy,Ellis:2023dgf, 2024PhRvD.109b3522E, 2024ApJ...976..212S, 2024ApJ...971L..10L, 2025JCAP...01..017B, 2025ApJ...978...31A, 2025PhRvD.111b3043S,2025PhRvD.111d3022X,Lamb:2025niq}. The brightest SMBHB in each frequency bin is expected to carry the leading non-Gaussianity and anisotropy  \cite{2013PhRvD..88h4001T, 2022ApJ...941..119B}. Even if this source is not individually detected as a continuous wave, it can still distort the inferred stochastic spectrum and leave measurable deviations in the population statistics. Current PTA searches have not yet found a confirmed individual SMBHB, but they already place upper limits on such sources \cite{2023ApJ...951L..50A, 2024A&A...690A.118E, Aggarwal:2018mgp,Becsy:2022zbu,NANOGrav:2023wsz, EPTA:2023gyr, Zhao:2025pgg, 2025MNRAS.536.1501G}. Therefore, the non-Gaussianity of the nHz background provides a promising way to test its astrophysical origin.

In this Letter, we introduce a two-stage hierarchical Bayesian framework for inferring the population properties of nHz GW sources. First, we perform a free-spectrum reconstruction \cite{2013PhRvD..87j4021L, 2014MNRAS.437.3004L, 2024PhRvD.109j3012J} in each frequency bin under two different hypotheses:  a Gaussian and isotropic GWB ($\mathcal{H}_0$); a GWB plus bright SMBHB sources ($\mathcal{H}_1$). 
Second, we use the free-spectrum posterior samples to construct a hierarchical likelihood for the GW source population with two different hypotheses: a large number of sources so that central limit theorem applies ($\mathcal{H}_{\rm inf}$) as speculated in various early Universe processes \cite{2023ApJ...951L..11A}; a finite number of SMBHBs ($\mathcal{H}_{\rm fin}$). This allows us to combine the information from the brightest source and the unresolved background consistently, and to infer both the GWB spectral properties and the underlying population parameters. By incorporating finite-source statistics, the framework extracts the non-Gaussian information encoded in the brightest binaries in each frequency band (step one) and in the power spectrum fluctuations across different frequency bands (step two), providing a direct Bayesian comparison between an astrophysical SMBHB population and a Gaussian random-field description of the GWB.

Throughout this work we use natural units with $G=c=1$. When cosmological quantities are required, we assume a flat $\Lambda$CDM cosmology with $\Omega_\mathrm{m}=0.3089$, $\Omega_\mathrm{b}=0.0486$, $\Omega_{\Lambda}=0.6911$, and $h=0.6774$ \citep{2020A&A...641A...6P}. 

\noindent{\textit{\textbf{GWB and the PTA Response.}}}
In the frequency domain, the correlation of the timing residual $R(f)$ is given by \cite{2021arXiv210513270T} 
\begin{equation}
    \langle R_a(f)R_b(f')\rangle=\frac{S_h(f)}{24\pi^2 f^2}\Gamma_{ab}\delta(f-f'),
\end{equation}
where $S_h(f)$ is the one-sided power spectral density (PSD) of the SGWB, $\delta(f-f')$ is the Dirac delta function, and
\begin{equation}
    \Gamma_{ab}=\frac{3}{2}x_{ab}\ln(x_{ab})-\frac{1}{4}x_{ab}+\frac{1}{2}+\frac{1}{2}\delta_{ab}
\label{eq:HD}
\end{equation}
is the Hellings \& Downs (HD) curve \cite{1983ApJ...265L..39H}. Here, $x_{ab}=(1-\cos\theta_{ab})/2$, and $\theta_{ab}$ denotes the angular separation between pulsars $a$ and $b$. The last term in Eq.~(\ref{eq:HD}) contributes only to the autocorrelation.


For a GWB produced by SMBHBs, the characteristic strain $h_{\rm c}(f)\equiv\sqrt{fS_h(f)}$ in the frequency bin $[f-1/2T,f+1/2T)$ can be expressed as the sum of the squared amplitudes of individual GW sources \cite{2004ApJ...611..623S}
\begin{equation}
\begin{split}
    h_{\rm c}^2(f) &=fT\left\langle h_+^2+h_\times^2\right\rangle_{\cos\iota,\psi,\phi} \\
    &=\frac{2}{5}fT\sum_j h_j^2,
\end{split}
\label{eq:hc}
\end{equation}
where $T$ is the observation period, $\left\langle\cdots\right\rangle_{\cos\iota,\psi,\phi}$ represents the average over inclination, polarization, and initial phase, $h_+$ and $h_\times$ are two polarization modes of GWs, and
\begin{equation}
    h_j = \frac{4\mathcal{M}_{c,j}^{5/3}(\pi f_j)^{2/3}}{d_{L,j}}
\end{equation} 
is the strain amplitude of the $j$-th binary, $\mathcal{M}_{c,j}$, $f_j$, $d_{L,j}$ are the chirp mass, frequency, luminosity distance of the $j$-th source, respectively. 
In the limit of an infinite number of SMBHBs, $h_{\rm c}(f)$ is simply  \cite{1995ApJ...446..543R, 2001astro.ph..8028P, 2003ApJ...583..616J, 2003ApJ...590..691W}
\begin{equation}
    h_{\rm c}(f)=A_\mathrm{GWB}\left (\frac{f}{f_\mathrm{yr}}\right)^{-2/3},
\label{eq:spectrum}
\end{equation}
where $A_\mathrm{GWB}$ is the GWB amplitude at the reference frequency $f_\mathrm{yr}=1\ \mathrm{yr}^{-1}$.

\noindent{\textit{\textbf{Hierarchical Bayesian Inference.}}}
To extract the population information of nHz GW sources, we apply a hierarchical Bayesian likelihood \cite{2019PASA...36...10T, 2019MNRAS.486.1086M, 2022hgwa.bookE..45V, 2023PhRvX..13a1048A},
 \begin{equation}
\mathcal{\tilde L}(\mathbf{R}|\mathbf{\Lambda})=\prod_i\int\mathcal{L}[\mathbf{R}(f_i)|\mathcal{H}, \vec{\theta}_i]p_{\rm pop}(\vec{\theta}_i|\tilde{\mathcal{H}},\mathbf{\Lambda})\dif\vec{\theta}_i\ . 
\label{eq:likelihood_total}
\end{equation}
where $\mathbf{R}(f_i)$ is the PTA data vector in frequency range $f \in [f_i-\frac{1}{2T},\, f_i+\frac{1}{2T})$.  In Eq.~(\ref{eq:likelihood_total}), two levels of hypotheses are involved. The hypothesis $\mathcal{H}$ specifies the signal model used in the free-spectrum reconstruction of each individual frequency bin, while $\vec{\theta}_i$ denotes the corresponding free-spectrum parameters in the $i$-th bin. The likelihood $\mathcal{L}[R(f_i)|\mathcal{H},\vec{\theta}_i]$ gives the probability of obtaining the data $R(f_i)$ under hypothesis $\mathcal{H}$ with parameters $\vec{\theta}_i$. From the Bayes’ theorem, the posterior of $\vec{\theta}_i$ is
\begin{equation}
    \mathcal{P}[\vec{\theta}_i|\mathbf{R}(f_i)]=\frac{\mathcal{L}[\mathbf{R}(f_i)|\mathcal{H}, \vec{\theta}_i]\pi(\vec{\theta}_i)}{Z_{\mathcal{H},i}},
    \label{eq:posterior}
\end{equation}
where $\pi(\vec{\theta}_i)$ is the prior imposed, $Z_{\mathcal{H},i}$ is the Bayesian evidence for frequency bin $i$.

The population hypothesis $\tilde{\mathcal{H}}$  with  parameters $\mathbf{\Lambda}$ specifies the probability density function (PDF) of $\vec{\theta}_i$,  $p_{\rm pop}(\vec{\theta}_{i}|\tilde{\mathcal{H}})$. With the free-spectrum posteriors, the population likelihood in Eq.~(\ref{eq:likelihood_total}) can be approximated as 
\begin{equation}
\begin{split}
    \tilde{\mathcal{L}}(\mathbf{R}|\mathbf{\Lambda}) \approx\prod_i\frac{1}{N_i}\sum_{ \substack{\vec{\theta}_{i,j}\sim \mathcal{P}(\vec{\theta}_i| \mathbf{R}(f_i)) \\ j=1, \cdots, N_{i}}}\frac{p_{\rm pop}(\vec{\theta}_{i,j}|\tilde{\mathcal{H}}, \mathbf{\Lambda})}{{\pi}(\vec{\theta}_{i,j})}\ ,
\end{split}
\label{eq:likelihood_hierarchical}
\end{equation}
where $N_i$ is the number of MCMC posterior samples obtained in the free-spectrum reconstruction, ${\pi}(\vec{\theta}_{i,j})$ is the prior used in the free spectrum reconstruction. And the posterior of $\mathbf{\Lambda}$ is
\begin{equation}
    \mathcal{P}[\mathbf{\Lambda}|\mathbf{R}]=\frac{\tilde{\mathcal{L}}(\mathbf{R}|\mathbf{\Lambda})\tilde{\pi}(\mathbf{\Lambda})}{\tilde{Z}_{\tilde{\mathcal{H}}} },
\label{eq:posterior_pop}
\end{equation}
where $\tilde{\pi}(\mathbf{\Lambda})$ is the population model prior, $\tilde{Z}_{\tilde{\mathcal{H}}}$ is the Bayesian evidence for the hierarchical inference.

The log Bayes factor,  $\Delta \ln Z^A_B = \ln Z_A-\ln Z_B$, quantifies the relative evidence provided by the data in favor of model A over  B. For the population statistics we are 
investigating, 
\begin{equation}\label{eq:lnZ_12}
    \Delta \ln Z^A_{B} =  \underbrace{\ln\frac{\prod_i Z_{A,i}}{\prod_i Z_{B,i}}}_{\Delta_{(1)}\ln Z^A_B} + \underbrace{\ln\frac{\prod_i \tilde Z_{A,i}}{\prod_i \tilde Z_{B,i}}}_{\Delta_{(2)}\ln Z^A_B}\ ,
\end{equation}
where $\Delta_{(1)}\ln Z$ and $\Delta_{(2)}\ln Z$ are contributed by free spectrum reconstruction and population inference, respectively (see Supplemental Material for derivation details).

\textit{\textbf{(1) Free Spectrum Reconstruction $\mathcal{H} = \mathcal{H}_{0/1}$.}} Assuming a Gaussian and isotropic background ($\mathcal{H}_0$), the free spectrum likelihood is 
\begin{equation}
    \log \mathcal{L}[\mathbf{R}(f_i)|\mathcal{H}_0, \vec{\theta}_i]=-\mathbf{R}^*(f_i)\mathbf{C}_i^{-1}\mathbf{R}(f_i)-\frac{1}{2}\log[\pi\det(\mathbf{C}_i)]\ ,
\label{eq:likelihood}
\end{equation}
where \cite{2023ApJ...951L..50A, 2024A&A...690A.118E, Aggarwal:2018mgp,Becsy:2022zbu,NANOGrav:2023wsz, EPTA:2023gyr, Zhao:2025pgg}
\begin{equation}
    C_{ab}(f)\equiv\left[\frac{h_{\rm c}^2(f)}{12\pi^2 f^3}\Gamma_{ab}+ (S_\mathrm{CURN}+S_\mathrm{white})\delta_{ab}\right]\frac{1}{T},
\end{equation}
is a $N_\mathrm{psr}\times N_\mathrm{psr}$ covariance matrix, and $\vec{\theta}_i = \{h_{{\rm c}, i}\}$.
Here $N_\mathrm{psr}$ is the number of pulsars,  $S_\mathrm{CURN}$ and $S_\mathrm{white}$ represent the PSD of the spatially-uncorrelated common red noise (CURN) and white noise, respectively.
 
When bright sources are present, the Gaussian and isotropic assumption may be violated. 
After subtracting a number of bright sources, the remaining signal can still be approximated as a Gaussian and isotropic background 
as shown in previous studies \cite{2013PhRvD..88h4001T,2022ApJ...941..119B}. In this case, the likelihood is formulated  as
\begin{equation}
\begin{split}
    \log \mathcal{L}[\mathbf{R}(f_i)|\mathcal{H}_1, \vec{\theta}_i]=& -\frac{1}{2}\log[\pi\det(\mathbf{C}_i)] \\
    -\left[\mathbf{R}^*(f_i)-\sum_{j}\mathbf{s}^*(\vec{r_j})\right] & \mathbf{C}_i^{-1}\left[\mathbf{R}(f_i)
    -\sum_{j}\mathbf{s}(\vec{r_j})\right] \ , 
\end{split}
\label{eq:likelihood2}
\end{equation}
where $\mathbf{s}(\vec{r_j})$ is the PTA response of the $j$-th bright SMBHB in the frequency bin, and $\vec{r_j}$ denotes the model parameters. The likelihood in Eq.~(\ref{eq:likelihood2}) is commonly used in search for individual binaries in PTA data \cite{2023ApJ...951L..50A, 2024A&A...690A.118E, Aggarwal:2018mgp,Becsy:2022zbu,NANOGrav:2023wsz, EPTA:2023gyr, Zhao:2025pgg}. Due to computational limitations, we only explicitly model the brightest SMBHB in each frequency bin, with parameters $\vec{r}_0\equiv\{h_0,\ f_0,\ \alpha,\ \sin\delta,\ \cos\iota,\ \psi,\ \phi\}$, and absorb the remaining binaries into an unresolved stochastic component. Here $h_0$ is the GW strain amplitude, $f_0$ is the GW frequency, $\alpha$  and $\delta$ denote the source sky location (see more details in Supplemental Material). In this case, $\vec{\theta}_i= \{ h_{\rm c}, {\vec r}_0\}_i$, where $h_{\rm c}$ is the characteristic strain of the unresolved component.


\textit{\textbf{(2) Population Inference $\tilde{\mathcal{H}} = \mathcal{H}_{\rm inf/fin}$}}.  
We start
with a simple hypothesis $\mathcal{H}_{\rm inf}$ where a large number of sources is assumed so that the central limit theorem applies,
and the PDF is simply 
\begin{equation}\label{eq:p_pop_H1}
    p_{\rm pop}(\vec{\theta_i}|\mathcal{H}_{\rm inf}, \mathbf{\Lambda})=\delta[h_{c,i}-h_{\rm c}(f_i;\mathbf{\Lambda})],
\end{equation}
where $\delta(x)$ is the Dirac delta function, and
\begin{equation}
    h_{\rm c}(f_i; \mathbf{\Lambda}) =A_\mathrm{GWB}\left(\frac{f_i}{f_\mathrm{yr}}\right)^{(-\gamma_{\mathrm{HD}}+3)/{2}}\ ,
    \label{eq:hc_power_law}
\end{equation}
with $\mathbf{\Lambda}=\{\gamma_\mathrm{HD},A_\mathrm{GWB}\}$.  The power-law power spectrum $\mathcal{H}_{\rm inf}$, together with the isotropic Gaussian field hypothesis $\mathcal{H}_0$ has been widely used for describing nHz GWs from various early universe processes \cite{2023ApJ...951L..11A}.

For astrophysical SMBHBs, a more realistic hypothesis $\mathcal{H}_{\rm fin}$ is the finite number of binaries. The PSD of GWB in each frequency bin therefore fluctuates, and these fluctuations become more significant at high frequencies where less binaries reside  \cite{2008MNRAS.390..192S, 2022ApJ...941..119B}.  The method of calculating the population PDF $p_{\rm pop}(\vec{\theta_i}|\mathcal{H}_{\rm fin}, \mathbf{\Lambda})$ in each frequency bin $f_i$ for a SMBHB population model has been proposed in \cite{Ellis:2023owy,2025PhRvD.111b3043S,2025PhRvD.111d3022X}. Since the PDF of source parameters $(\alpha,\ \sin\delta,\ \cos\iota,\ \psi,\ \phi)$ is independent of the SMBHB population model and is uncorrelated with $h_0$ and $h_{\rm c}$,   the PDF of $\vec{\theta}_i$ can be formulated as 
\begin{equation}
    p_{\rm pop}(\vec{\theta}_i|\mathcal{H}_{\rm fin}, \mathbf{\Lambda}) = {\rm const} \times p_{\rm pop}(S_i|\mathcal{H}_{\rm fin}, \mathbf{\Lambda})\ ,
\end{equation} 
where $S_i = \sum_s (h^2(f_i))_s$ with $s$ being the label of SMBHBs in frequency bin $f_i$. 
For the hypothesis $\mathcal{H}_1$ of GWB+bright sources, 
\begin{equation}
    S_i=\frac{5h_{c,i}^2}{2f_i T}+h_{0,i}^2\ .
    \label{eq:S}
\end{equation}

As a proof of principle, we consider a simple SMBHB population model. 
In addition to the expected value of characteristic strain $h_{\rm c}(f_i)$ of the GWB (Eq.~\ref{eq:hc_power_law}),  
the strain amplitudes $h$ of SMBHBs follows a power-law distribution 
\begin{equation}
    g(h^2)=A h^{-2 \gamma},\quad h_\mathrm{min}\leq h<h_\mathrm{max},
\label{eq:fs}
\end{equation}
where $\gamma$ is the power-law index and $A$ is the normalization constant such that $\int_{h_\mathrm{min}^2}^{h_\mathrm{max}^2}g(h^2) d h^2 =1 $.  The expected source number $\bar{N}$ in each frequency bin is 
\begin{equation}
    \bar{N}=\frac{5h_{\rm c}^2}{2fT}\left[\int_{h_{\rm min}^2}^{h_{\rm max}^2} g(h^2)h^2\dif h^2\right]^{-1}.
\end{equation}



With the population information encoded in $g(h^2)$ and $\bar N$, the PDF can be formulated as  \cite{Ellis:2023owy,2025PhRvD.111b3043S,2025PhRvD.111d3022X}
\begin{equation}
    p_{\rm pop}(S|\mathcal{H}_{\rm fin}, \mathbf{\Lambda}) = \frac{1}{\pi}\mathrm{Re}\left\{ \int_0^\infty e^{\bar{N}[\hat{g}(\omega)-1] +i\omega S} \dif \omega\right\},
\label{eq:PS}
\end{equation}
where $\mathrm{Re}$ denotes the real part, $\hat{g}(\omega)$ is the Fourier transform of $g(h^2)$ and $\mathbf{\Lambda}=\{A_{\rm GWB},\ \gamma_{\rm HD},\ \gamma\}$. Eq.~(\ref{eq:PS}) provides the statistical connection between the population hypothesis $\mathcal{H}_{\rm fin}$  and the observed  fluctuations in $S$. 
To simplify the calculation and reduce the computational cost, we show the inference results with  fixed $h_{\rm min}$ and $h_{\rm max}$ (in Supplemental Material, we show the inference without fixing the cutoff amplitudes, which yields consistent results).





\noindent{\textit{\textbf{Simulations.}}} We simulate a PTA with $N_\mathrm{psr}=200$ pulsars with an observation period of $T=30\ {\rm yr}$, and assume a white noise with standard deviation $\sigma_n=200$ ns in TOAs of each pulsar. For CURN, we assume that its one-sided PSD has the form
\begin{equation}
    S_\mathrm{CURN}(f)=\frac{A^2_\mathrm{CURN}}{12\pi^2}\left(\frac{f}{f_\mathrm{yr}}\right)^{-\gamma_\mathrm{CURN}}f_\mathrm{yr}^{-3}.
\end{equation}
Here we adopt the reference amplitude and power index as $A_\mathrm{CURN}=1\times 10^{-15}$ and $\gamma_\mathrm{CURN}=4.0$. 

We generate mock data of GWs from the population model described by Eqs.~(\ref{eq:hc_power_law},\ref{eq:fs}), with $A_{\mathrm{GWB}}=2.4\times10^{-15}$ and $\gamma_\mathrm{HD}=13/3$, which satisfy the constraints of NANOGrav 15-year data \cite{2023ApJ...951L...8A}. We choose fiducial values of $h_{\mathrm{min/max}}$ as 
$h_{\mathrm{min}}(f_\mathrm{yr})=2\times10^{-16}$, $h_{\mathrm{max}}(f_\mathrm{yr})=2\times 10^3 h_{\mathrm{min}}(f_\mathrm{yr})$. The lower limit ensures that most of the GWB energy predicted by previous semi-analythic models is included \cite{1995ApJ...446..543R, 2003ApJ...583..616J, 2003ApJ...590..691W, 2004ApJ...615...19E, 2008MNRAS.390..192S, 2009MNRAS.394.2255S, 2013MNRAS.433L...1S, 2020ApJ...897...86C, 2023ApJ...949L..24S, 2023ApJ...952L..37A, 2026JCAP...02..001L}, while the upper limit guarantees that the simulated GW amplitudes remain below current observational constraints \cite{2023ApJ...951L..50A, 2024A&A...690A.118E}. For other frequency bands, the lower and upper bounds are naturally 
$h_{\mathrm{min/max}}(f_i)=\left(f_i/f_\mathrm{yr}\right)^{2/3}h_\mathrm{min/max}(f_\mathrm{yr})$, assuming the frequency evolution of SMBHBs is governed by GW emissions.

\noindent{\textit{\textbf{Inference.}}} In the main text, we focus on the simulation and inference with $\gamma=2.2$, and show the comparison with the case with $\gamma=2.5$ in Supplemental Material \footnote{In the high-amplitude tail, Ref.~\cite{Raidal:2026ezm} shows the universal behavior, $g(h^2) \rightarrow (h^2)^{-5/2}$, independent of population models. We use a universal power law index that may be different from $2.5$ for taking the low-amplitude regime into account.}.

\begin{figure}[h]
    \centering
   \includegraphics[width=8.5cm]{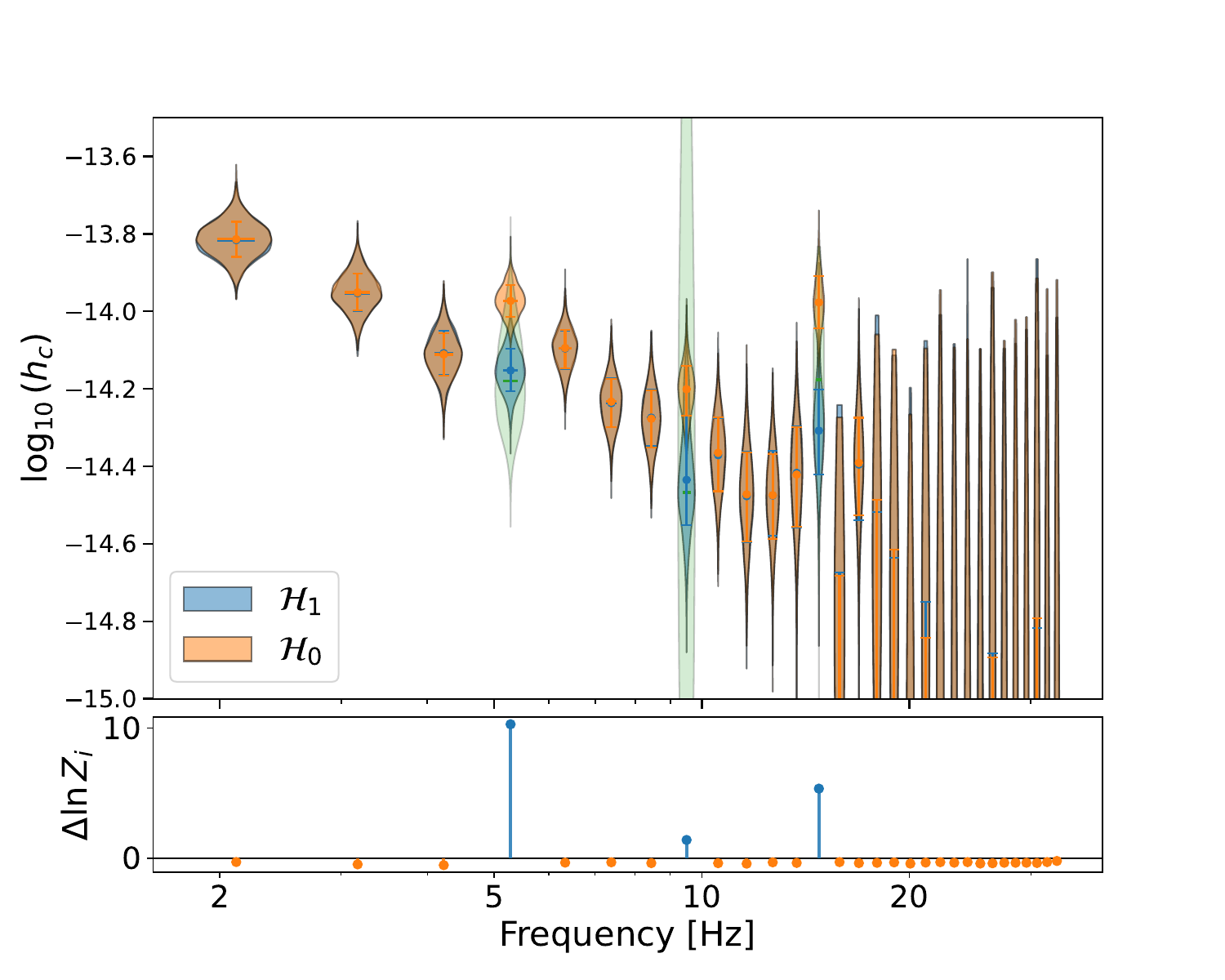}
    \caption{The upper panel presents the posterior distributions of $h_{c}$ in 30 frequency bins, with orange violins for the hypothesis $\mathcal{H}_0$ (GWB only) and blue violins for the $\mathcal{H}_1$ (GWB+bright sources).  The lower panel shows the log Bayes factor in each bin and the total log Bayes factor is $ \Delta_{(1)} \ln Z_{\mathcal{H}_0}^{\mathcal{H}_1} = 7.23$. Posteriors of $h_0\times(2fT/5)^{-{1/2}}$ in frequencies with a positive log Bayes factor are also plotted in green violins.} 
    \label{fig:spectrum}
\end{figure}

{\textit{\textbf{(1)   $\mathcal{H}_0$ (GWB only) versus $\mathcal{H}_1$ (GWB + bright sources).}}} 
To simplify the calculations,  we focus on inference of source parameters and assume that the noise parameters have been well measured. 
For each frequency band, we use the \texttt{bilby} \cite{2019ApJS..241...27A} package and \texttt{dynesty} \cite{2020MNRAS.493.3132S} sampler to explore the parameter space, with 2000 live points and a threshold of 10000 maximum MCMC steps.

Fig.~\ref{fig:spectrum} shows the free spectrum reconstruction for $\gamma=2.2$. 
Nonzero $h_{\rm c}$ is favored  in the lowest $\sim 14$ frequency bins, whereas
the higher-frequency bins mainly yield upper limits where the white noise dominates over the signal.
In this simulation,  a small number of SMBHBs can  dominate individual frequency bins and produce significant deviations from a power-law spectrum. Once the brightest source is modeled explicitly, these deviations are absorbed by the deterministic source component, and the power spectrum of the remaining unresolved background becomes smoother.

The lower panel of Fig.~\ref{fig:spectrum} shows the bin-wise log Bayes factors between hypotheses $\mathcal{H}_1$ and $\mathcal{H}_0$. From this simulation, the preference for $\mathcal{H}_1$ is found at $f\simeq 5\,{\rm nHz}, 9\,{\rm nHz}$ and $14\,{\rm nHz}$. Most other bins are inconclusive or slightly favor the $\mathcal{H}_0$ hypothesis. In the upper panel, we use the green violin plots to show the $h_0$ posteriors for bins where  $\mathcal{H}_1$ is favored. A larger Bayes factor indicates stronger evidence for an individual source. The successful identification of several individual SMBHBs from the background around 10 nHz is also consistent with recent simulation-based predictions \cite{2024ApJ...974..261C, 2025CQGra..42b5021C}.



\begin{figure}
    \centering
   \includegraphics[width=8.5cm]{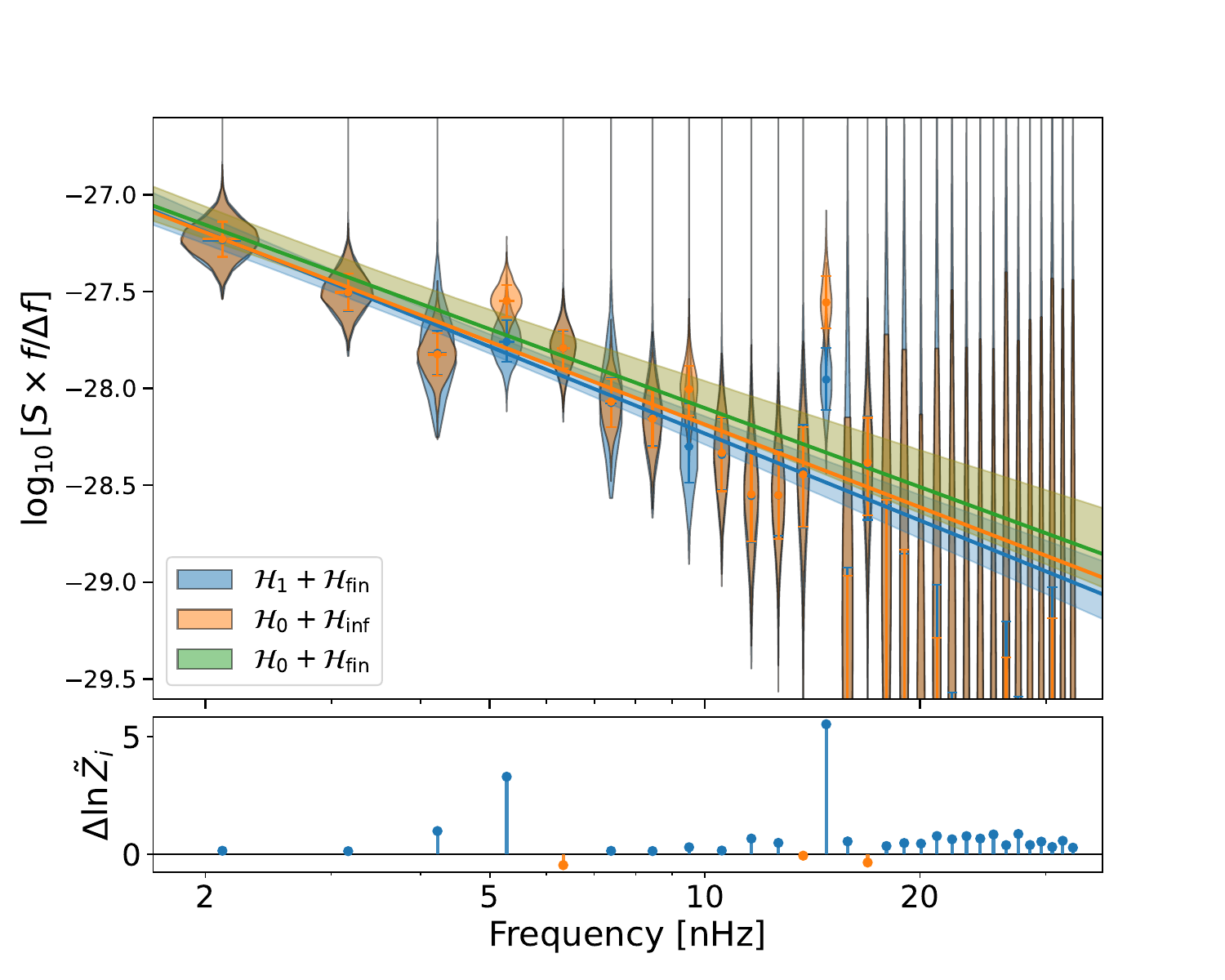}
    \caption{Hierarchical Bayesian inference results. The solid curves are the best-fit spectra under three different hypotheses, and the shaded bands indicate the corresponding $1\sigma$ credible intervals. The lower panel shows the log Bayes factor in each bin and the total contribution is $\Delta_{(2)} \ln Z_{\mathcal{H}_0+\mathcal{H}_{\rm inf}}^{\mathcal{H}_1+\mathcal{H}_{\rm fin}}  =18.79$ }
    \label{fig:pop}
\end{figure}

\textit{\textbf{(2) $\mathcal{H}_{\rm inf}$ (infinite number of sources) versus $\mathcal{H}_{\rm fin}$ (finite number of sources).}}
We then compare the inferred power $S(f_i)$ with the population models in Fig.~\ref{fig:pop}. 
Hypothesis $\mathcal{H}_{\rm inf}$ predicts a strict power spectrum in Eqs.~(\ref{eq:p_pop_H1},\ref{eq:hc_power_law}), which roughly fits the overall trend of $S(f_i)$ but fails to describe its bin-to-bin fluctuations.
These fluctuations are expected for a finite population of SMBHBs, and therefore can be better captured by the 
finite population hypothesis $\mathcal{H}_{\rm fin}$. The quantitative bin-wise preference $\Delta\ln \tilde{Z}$  for $\mathcal{H}_{\rm fin}$ is shown in the lower panel of Fig.~\ref{fig:pop}. Combining the free-spectrum reconstruction and the hierarchical population inference, the preference for $\mathcal{H}_1+\mathcal{H}_{\rm fin}$ over $\mathcal{H}_0+\mathcal{H}_{\rm inf}$ reaches $\Delta\ln Z=22.74$ for the 15 lowest frequency bins and $\Delta\ln Z=26.02$ for all the 30 frequency bins (see Table~\ref{tab:bf_gamma22}). This comparison shows that the data strongly prefer a finite SMBHB population over a Gaussian  isotropic background with a power-law spectrum.

\begin{table}[]
    \centering
    \begin{tabular}{ccc}
    \hline
        Model &  15 bins  & 30  bins \\
        \hline
        $\mathcal{H}_0+\mathcal{H}_{\rm inf}$ & -&-\\
       $\mathcal{H}_0+\mathcal{H}_{\rm fin}$ & $3.75$ & $8.79$\\
       $\mathcal{H}_1+\mathcal{H}_{\rm fin}$ & $22.74(12.45, 10.28)$ & $26.02(7.23, 18.79)$\\
        \hline
    \end{tabular}
    \caption{The preference for different hypotheses by the mock data with $\gamma=2.2$ is quantified with the Bayes factor $\Delta \ln Z (\Delta_{(1)} \ln Z, \Delta_{(2)} \ln Z) $,
    where the inference is conducted for 15 lowest frequency bins and all 30  bins, respectively. }
    \label{tab:bf_gamma22}
\end{table}

\begin{figure}
    \centering
   \includegraphics[width=8.5cm]{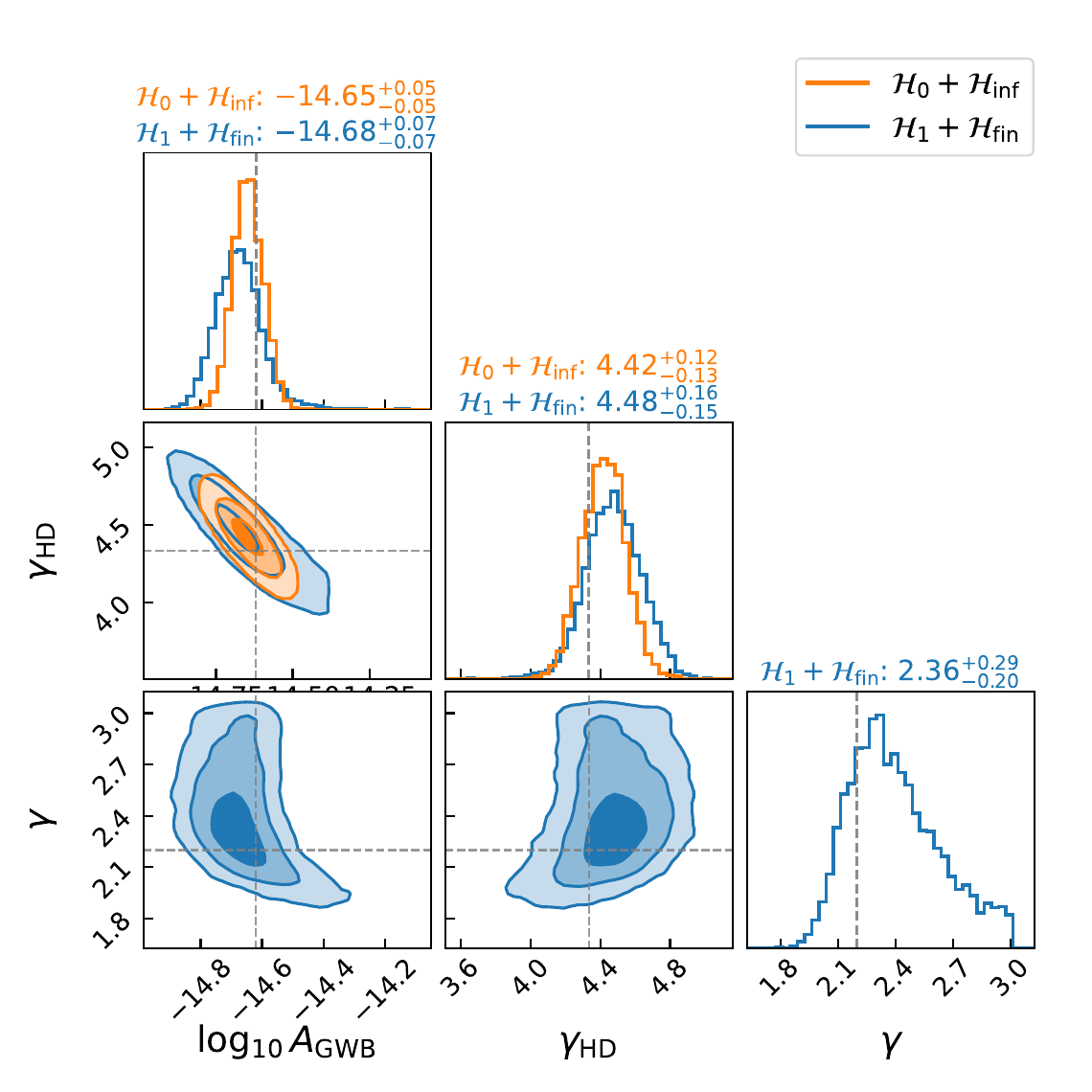}
    \caption{The corner plots for $\mathcal{H}_0+\mathcal{H}_{\rm inf}$ v.s. $\mathcal{H}_1+\mathcal{H}_{\rm fin}$. The values shown above the one-dimensional posterior panels indicate the median and the central $68\%$ credible interval for each parameter. }
    \label{fig:corner}
\end{figure}

In Fig.~\ref{fig:corner}, we show the posteriors of the population model parameters $\mathbf{\Lambda}$ in the corner plot. 
Consistent with Fig.~\ref{fig:pop}, the favored hypothesis $\mathcal{H}_1+\mathcal{H}_{\rm fin}$ correctly recovers the injection values of model parameters $A_{\rm GWB}, \gamma_{\rm HD}, \gamma$. The less favored hypothesis $\mathcal{H}_0+\mathcal{H}_{\rm inf}$ also recovers the injection values of $A_{\rm GWB}, \gamma_{\rm HD}$, since it captures the overall trend of the PSD.

As a comparison, we also analyze a mixture hypothesis $\mathcal{H}_0+\mathcal{H}_{\rm fin}$: a Gaussian and isotropic GWB is assumed in the free spectrum reconstruction, and a finite number of sources is assumed in the hierachical population inference. The mixture hypothesis was used in \cite{Ellis:2023dgf, 2025PhRvD.111b3043S}, and was numerically verified in \cite{2025PhRvD.111d3022X,Raidal:2026ezm}.
It is of no surprise to find the mixture hypothesis is more favored than $\mathcal{H}_0+\mathcal{H}_{\rm inf}$, and is less favored than 
$\mathcal{H}_1+\mathcal{H}_{\rm fin}$ by the mock data with $\gamma=2.2$. The preference ranking  would be different for 
a mildly non-Gaussian GWB from a finite population of SMBHBs (see  Supplemental Material for the inference result of mock data with $\gamma=2.5$).

\noindent{\textit{\textbf{Conclusion.}}}
In this Letter, we introduced a hierarchical inference framework for inferring population properties of nHz GW sources, 
by optimally leveraging non-Gaussianity (and anisotropy) of individual bright sources and in fluctuations of the PSD of the GWB. The hierarchical inference consists of two steps.
In the first step, the free-spectrum reconstruction is applied to PTA data under either the GWB-only hypothesis ($\mathcal{H}_0$) or the GWB+bright sources hypothesis ($\mathcal{H}_1$). In the second step, the free-spectrum reconstruction results are used for constraining different GW source population models, including $\mathcal{H}_{\rm inf}$ (infinite number of sources) and  $\mathcal{H}_{\rm fin}$ (finite number of sources).

As an example, we applied the hierarchical inference framework to  mock PTA data generated from a SMBHB population model, and we calculated the evidence for hypotheses $\mathcal{H}_0$ and $\mathcal{H}_1$ in each frequency bin. For a strongly non-Gaussian GWB ($\gamma=2.2$),  the free-spectrum reconstruction shows support for $\mathcal{H}_1$ in several frequency bins, i.e., the presence of an individual source in addition to a GWB. Along with the free-spectrum reconstruction, the hierarchical inference not only shows strong support for $\mathcal{H}_{\rm fin}$, but also constrains the population properties of SMBHBs.
For a mildly non-Gaussian GWB ($\gamma=2.5$), although the free-spectrum reconstruction shows support for $\mathcal{H}_0$ instead,
the hierarchical inference still shows mild support for $\mathcal{H}_{\rm fin}$ and yields a reasonable constraint on the population properties of SMBHBs (see Supplemental Material).

To summarize, the hierarchical Bayesian inference framework provides a practical way for optimally using non-Gaussianity (and anisotropy) in the nHz GWs. By combining the unresolved stochastic component with the brightest SMBHBs, it can help distinguish a GWB generated by astrophysical SMBHBs from  a Gaussian and isotropic GWB expected from various early-universe processes. Future applications to real PTA data, together with more realistic SMBHB population models or better summary statistics of the population models \cite{Raidal:2026ezm, Ali-Haimoud:2026sbk,Goncharov:2026cdh} and multiple bright-source components, are expected to place stronger constraints on the origin of the nHz GWB and on the population properties of SMBHBs.

~\\

{\it Note added:} As we were finishing this work, Ref.~\cite{Goncharov:2026cdh} was posted online, where a similar idea of hierarchical Bayesian inference with GWB+bright sources decomposition was proposed for inferring population properties of SMBHBs. In \cite{Goncharov:2026cdh}, the characteristic number of SMBHB sources, $N_{\rm c}$, was used as a detection statistic for the astrophysical origin of the GWB. In our work, we derived the Bayes evidence for hierarchical inference and used the Bayes factor $\Delta_{(1)}\ln Z + \Delta_{(2)}\ln Z$ for distinguishing different origins of the GWB.



~\\

\noindent{\textit{\textbf{Acknowledgements.}}}
The results in this paper have been derived using the following packages: \texttt{astropy}\cite{2022ApJ...935..167A}, \texttt{Bilby} \cite{2019ApJS..241...27A}, \texttt{corner} \cite{corner}, \texttt{dynesty} \cite{2020MNRAS.493.3132S}, \texttt{Jupyter} \cite{2016ppap.book...87K}, \texttt{numpy} \cite{2020Natur.585..357H}, \texttt{matplotlib} \cite{2007CSE.....9...90H}, \texttt{mpmath} \cite{mpmath}, \texttt{scipy} \cite{2020NatMe..17..261V}. This work made use of the Gravity Supercomputer at the Department of Astronomy, Shanghai Jiao Tong University. This work was supported by the National Natural Science Foundation of China (Grant No. 12505077). X.X.~is funded by the grant CNS2023-143767. 
Grant CNS2023-143767 is funded by MICIU/AEI/10.13039/501100011033 and by European Union NextGenerationEU/PRTR.

\bibliographystyle{apsrev}
\bibliography{main}

\clearpage
\onecolumngrid
\newpage
\begin{center}
  \textbf{\large{Supplemental Material}} \\
\end{center}
\twocolumngrid

\section{SMBHB Waveform}
\label{sec:waveform}
In general relativity, the GW tensor can be written as 
\begin{equation}
    h_{ij}(t)=h_+(t)\epsilon_{ij}^++h_\times(t)\epsilon^\times_{ij},
    \label{eq:hij}
\end{equation}
where $\epsilon_{ij}^+$ and $\epsilon^\times_{ij}$ are two polarization tensors, and for a circular SMBHB, the waveform can be written as \cite{Maggiore:2007ulw}
\begin{equation}
\begin{aligned}
    h_+(t)=h[&\cos\iota\sin2\psi\sin(2\pi f t+\phi)\\
    &-\frac{1}{2}(1+\cos^2\iota)\cos2\psi\cos(2\pi f t + \phi)],
    \label{eq_h1}
\end{aligned}
\end{equation}
\begin{equation}
\begin{aligned}
    h_\times(t)=-h[&\cos\iota\cos2\psi\sin(2\pi f t + \phi)\\
    &+\frac{1}{2}(1+\cos^2\iota)\sin2\psi\cos(2\pi f t + \phi)],
    \label{eq_h2}
\end{aligned}
\end{equation}
where $\iota$ is the inclination angle, $\psi$ is the polarization angle, $\phi$ is the initial phase, and the amplitude $h$ is
\begin{equation}
    h=\frac{4\mathcal{M}_c^{5/3}(\pi f)^{2/3}}{d_L},
    \label{eq:h0}
\end{equation}
where $\mathcal{M}_c$ is the redshift chirp mass, $d_L$ is the luminosity distance. 

When binary's separation falls below $\sim$1 parsec, the orbital evolution would be driven by GW emission, and the observed GW frequency evolves as \cite{1963PhRv..131..435P, 1964PhRv..136.1224P, Maggiore:2007ulw}
\begin{equation}
    \dot{f}=\frac{96}{5}\pi^{8/3}\mathcal{M}_c^{5/3}f^{11/3}.
\end{equation}
The time to coalescence from frequency $f$ is
\begin{equation}
    t_c-t=\frac{5}{256}\mathcal{M}_c^{-5/3}(\pi f)^{-8/3}.
\end{equation}
Usually, the evolution timescale is much longer than the observation period in the PTA band, so an individual SMBHB can be treated as nearly monochromatic within a single frequency bin. Therefore, we can obtain $\dif N/\dif f\propto f^{-8/3}$ and $h_{\rm c}(f)\propto f^{-2/3}$.

\section{Response to the brightest SMBHB}
Denote the parameters of the brightest SMBHB as $\vec{r}_0\equiv\{h_0,\ f_0,\ \alpha,\ \sin\delta,\ \cos\iota,\ \psi,\ \phi\}$, its waveform is described by the Eqs. (\ref{eq:hij})--(\ref{eq_h2}). For pulses propagating along the unit vector $\hat{p}$ with an initial frequency of $\nu_0$, their response to this brightest SMBHB can be written as \cite{1975GReGr...6..439E, 1979ApJ...234.1100D, 1987GReGr..19.1101W, 2010PhRvD..81j4008S,  Niu:2018oox, 2025PhRvD.112b3012Y}
\begin{equation}
\begin{split}
    s(t,\ \vec{r}_0)&\equiv\int_0^t \dif t'\frac{\nu(t') - \nu_0}{\nu_0}\\
    &=\int_0^{t}\dif t'\sum_{A=+,\times}\left[ F^A\Delta h_A(t',\ \vec{r}_0)\right],
\label{eq:st}
\end{split}
\end{equation}
where
\begin{equation}
    F^{A}\equiv\frac{\hat{p}_i\hat{p}_j}{2(1+\hat{\Omega}\cdot\hat{p})}\epsilon^A_{ij}(\hat{\Omega}),\quad A=+,\times,
\end{equation}
are the antenna beam patterns, $\hat{\Omega}=-(\cos\delta\cos\alpha,\ \cos\delta\sin\alpha,\ \sin\delta)$ is the GW propagation direction, $\Delta h_A(t,\ \vec{r}_0)$ is the difference of waveform between the pulsar and the earth, which are named as the pulsar term and the earth term, respectively. Generally, in actual SMBHB searches, ignoring the pulsar term may weaken the parameter constraints and introduce parameter biases \cite{2010arXiv1008.1782C, 2016MNRAS.461.1317Z}. However, including the pulsar term would introduce additional pulsar-dependent phases, and possibly frequency-evolution effects, which depend on the poorly known pulsar distances. Since the main purpose of this study is to test whether a bright SMBHB source can be statistically separated from an unresolved background and then used in a hierarchical population inference, we neglect the pulsar term, both in the signal injection and in the recovery.

After neglecting the pulsar term, the response is
\begin{equation}
    s(t,\ \hat\Omega)=-\frac{1}{2f_0}\sum_{A=+,\times}F^{A} h_A(t,\ \vec{r}_0).
\end{equation}
In the frequency domain, it can be rewritten as
\begin{equation}
    s(f_i,\ \vec{r}_0)=-\frac{1}{2f_0}\sum_{A=+,\times}F^{A} \hat{h}_A(f_i|\vec{r}_0),
\end{equation}
where
\begin{equation}
    \hat{h}(f_i|\vec{r}_0)=\frac{1}{T}W(f_i-f_0)h_0\mathfrak{g}(\alpha,\delta,\iota,\psi)e^{i\phi}
\end{equation}
represents the Fourier transform of the function $h_0\mathfrak{g}(\alpha,\delta,\iota,\psi)e^{i(2\pi f_0t+\phi)}$, and $W(f_i-f_0)$ is the top-hat window function,
\begin{equation}
    W(f_i-f_0)\equiv\begin{cases}
        1,\quad f_i-1/2T<f_0\leq f_i+1/2T,\\
        0,\quad \mathrm{otherwise}.
    \end{cases}
\label{eq:window}
\end{equation}

\section{Evidence calculation in the Hierarchical Bayesian inference}

From Eq.~(\ref{eq:posterior}), the  evidence  for hypothesis $\mathcal{H}$ by data in $i$-th frequency bin is given by
\begin{equation}\label{eq:Z_H}
     Z_{\mathcal{H}}[\mathbf{R}(f_i)] = \int \mathcal{L}[\mathbf{R}(f_i)|\mathcal{H}, \vec{\theta}_i]\pi(\vec{\theta}_i) d\vec{\theta}_i\ ,
\end{equation} 
and the free-spectrum likelihood can be written as
\begin{equation}
    \mathcal{L}[\mathbf{R}(f_i)|\mathcal{H}, \vec{\theta}_i]=\frac{\mathcal{P}[\vec{\theta}_i|\mathbf{R}(f_i)]Z_{\mathcal{H},i}}{\pi(\vec{\theta}_i)}.
\end{equation}
Substitute it into Eq. (\ref{eq:likelihood_total}), we obtain the population likelihood
\begin{equation}
\begin{split}
    \mathcal{L}(\mathbf{R}|\mathbf{\Lambda})&=\prod_i\int \frac{\mathcal{P}[\vec{\theta}_i|\mathbf{R}(f_i)]Z_{\mathcal{H},i}}{\pi(\vec{\theta}_i)}p_{\rm pop}(\vec{\theta}_i|\tilde{\mathcal{H}}, \mathbf{\Lambda})\dif\vec{\theta}_i\\
    &\approx\prod_i\frac{Z_{\mathcal{H},i}}{N_i}\sum_{ \substack{\vec{\theta}_{i,j}\sim \mathcal{P}(\vec{\theta}_i| \mathbf{R}(f_i)) \\ j=1, \cdots, N_{i}}}\frac{p_{\rm pop}(\vec{\theta}_{i,j}|\tilde{\mathcal{H}}, \mathbf{\Lambda})}{{\pi}(\vec{\theta}_{i,j})}\ ,
\end{split}
\end{equation}
while the approximate equal sign is equivalent to performing a Monte Carlo integration using MCMC samples from the free spectrum reconstruction. In this step, the evidence $Z_{\mathcal{H},i}$ of the free-spectrum reconstruction is usually dropped, since it has no dependence on the population parameters $\mathbf{\Lambda}$. Then the likelihood above simplifies as $\mathcal{\tilde L}(\mathbf{R}|\mathbf{\Lambda})$ in Eq.~(\ref{eq:likelihood_hierarchical}). 
In this form, the total evidence can be clearly divided as two components as in Eq.~(\ref{eq:lnZ_12}), where the second component,  the evidence of the hierarchical inference $\tilde{\mathbf{Z}}_{\tilde{\mathcal{H}}}$,  can be calculated as 
\begin{equation}
\begin{split}
    \frac{1}{\tilde{\mathbf{Z}}_{\tilde{\mathcal{H}}}}=\int\frac{\mathcal{P}(\mathbf{\Lambda|\mathbf{R}})}{\tilde{\mathcal{L}}(\mathbf{R}|\tilde{\mathcal{H}}, \mathbf{\Lambda})}\dif\mathbf{\Lambda}
    \approx\frac{1}{N_{\rm H}}\sum_{j=1}^{N_\mathrm{H}} \frac{1}{\tilde{\mathcal{L}}[\mathbf{R}|\tilde{\mathcal{H}}, \Lambda_j]},
\end{split}
\end{equation}
while $N_\mathrm{H}$ is the number of MCMC posterior samples in the hierarchical inference. The contribution of each frequency bin
can be calculated as 
\begin{equation}
    (\tilde Z_{\tilde{\mathcal{H}},i})^{-1}\equiv\frac{1}{N_\mathrm{H}}\sum_{j=1}^{N_\mathrm{H}} \frac{1}{\tilde{\mathcal{L}}[\mathbf{R}(f_i)|\tilde{\mathcal{H}}, \Lambda_j]}\ .
\end{equation}
This bin-wise evidence estimate is then used to compute the per-bin log Bayes factor shown in the lower panel of Fig.~\ref{fig:pop}, by taking the log-evidence difference between the two population hypotheses in each frequency bin.




\section{More details of Simulations and inference }

The priors of model parameters used in the Bayesian inference in the main text are summarized in Table~\ref{tab:priors}.

\begin{table}[htbp]
    \centering
    \begin{tabular}{cccc}
    \hline\hline
        Parameter & Prior & Minimum & Maximum\\
        \hline
        $h_{\rm c}$ & LogUniform&$10^{-5}h_{\rm c}$&$10^{5}h_{\rm c}$\\
        $h_0$ & LogUniform&$10^{-5}h_{\rm 0}$&$10^{5}h_{\rm 0}$\\
        $\alpha$ & Uniform & 0 & $2\pi$\\
        $\sin\delta$ & Uniform & -1 & 1\\
        $\cos\iota$ & Uniform & 0 & 1\\
        $\psi$ & Uniform & 0 & $\pi$\\
        $\phi$ & Uniform & 0 & $2\pi$\\
        $f_0$ & Uniform & $f_0-\Delta f/2$ & $f_0+\Delta f/2$\\
        \hline
        $A_{\mathrm{GWB}}$&LogUniform&$1\times 10^{-18}$&$1\times 10^{-13}$\\
        $\gamma_\mathrm{HD}$&Uniform&0&10\\
        $\gamma$&Uniform&1.5&3\\
        \hline
    \end{tabular}
    \caption{{Injection values and priors of model parameters in simulations and inference. The first part shows the parameters in the free spectrum analysis, and the second part lists the parameters in the hierarchical Bayesian inference.} }
    \label{tab:priors}
\end{table}

\begin{figure}[h]
    \centering
   \includegraphics[width=8.cm]{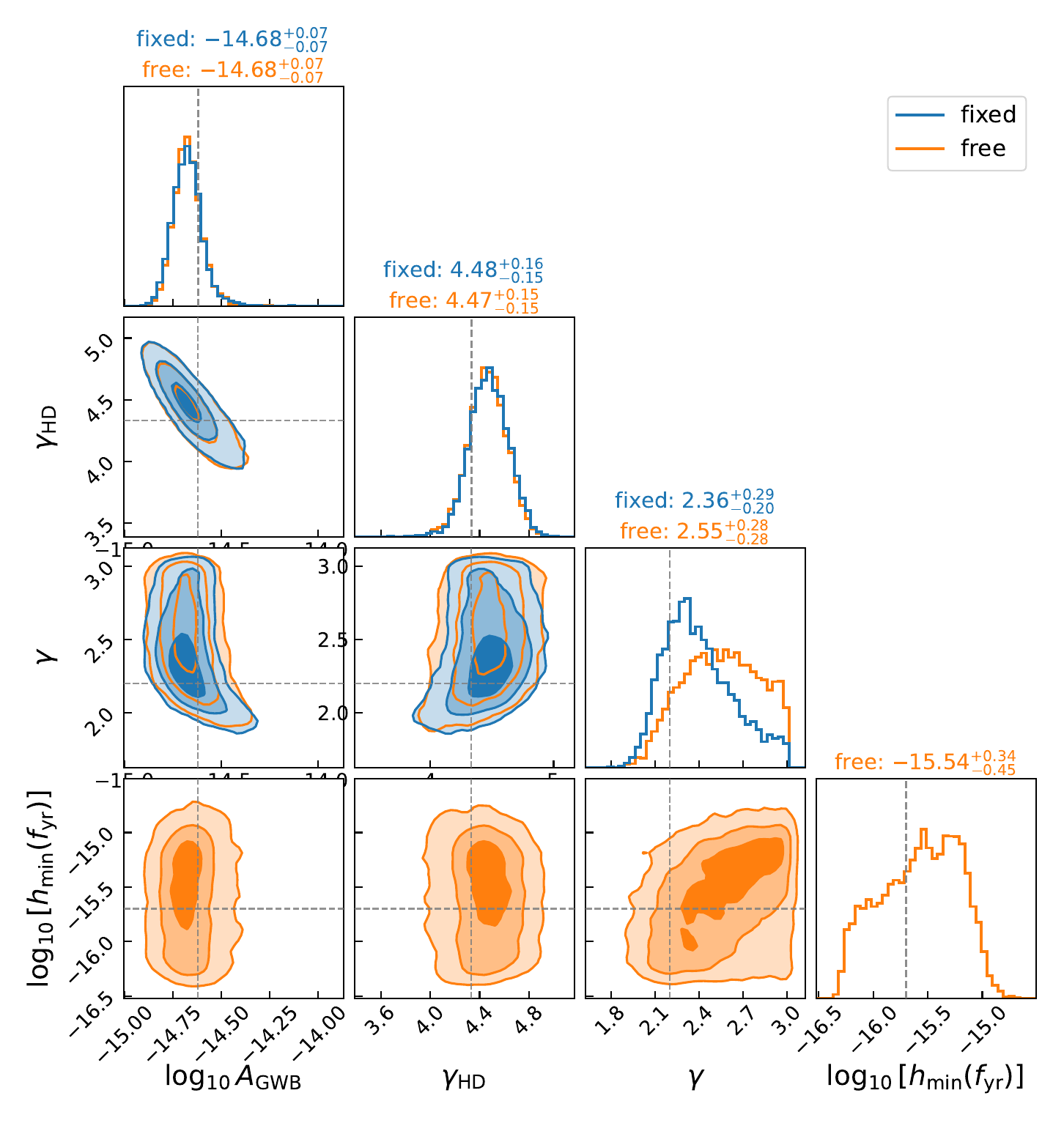}
    \caption{Comparison of constraints of population model parameters with fixed and free $h_{\min}(f_{\rm yr})$.}
    \label{fig:corner_3}
\end{figure}

As a consistency check, we conduct the Bayesian inference of the mock data in the main text without fixing the lower GW amplitude limit $h_{\rm min}(f_{\rm yr})$, and show the constraints of the population model parameters in Fig.~\ref{fig:corner_3}. Comparison with  Fig.~\ref{fig:corner} in the main text shows that fixing $h_{\rm min}(f_{\rm yr})$ makes little difference to the  GWB spectrum parameters $A_{\rm GWB}$ and $\gamma_{\rm HD}$, since they are constrained by the overall trend of the power spectrum of the GWB. The main impact is on the posterior of power index $\gamma$, due to its partial degeneracy with $h_{\rm min}(f_{\rm yr})$. 
In addition, Bayesian inference with a free lower limit  $h_{\rm min}(f_{\rm yr})$ leads a small variation in the evidence, $\Delta_{(2)} \ln Z_{\rm free}^{\rm fixed} =1.19$.

\begin{table}[]
    \centering
    \begin{tabular}{ccc}
    \hline
        Model & 15 bins & 30 bins\\
        \hline
        $\mathcal{H}_0+\mathcal{H}_{\rm inf}$ & -& -\\
        $\mathcal{H}_0+\mathcal{H}_{\rm fin}$ & -0.02 & 2.45\\
        $\mathcal{H}_1+\mathcal{H}_{\rm fin}$ & $-3.11(-5.19+2.07)$ & $-5.32(-13.79+8.47)$\\
        \hline
    \end{tabular}
    \caption{Same as Table~\ref{tab:bf_gamma22}, but for $\gamma=2.5$.}
    \label{tab:bf_gamma25}
\end{table}

\section{Bayesian Inference for mock data with $\gamma=2.5$} 

\begin{figure}[h]
    \centering
   \includegraphics[width=8.0cm]{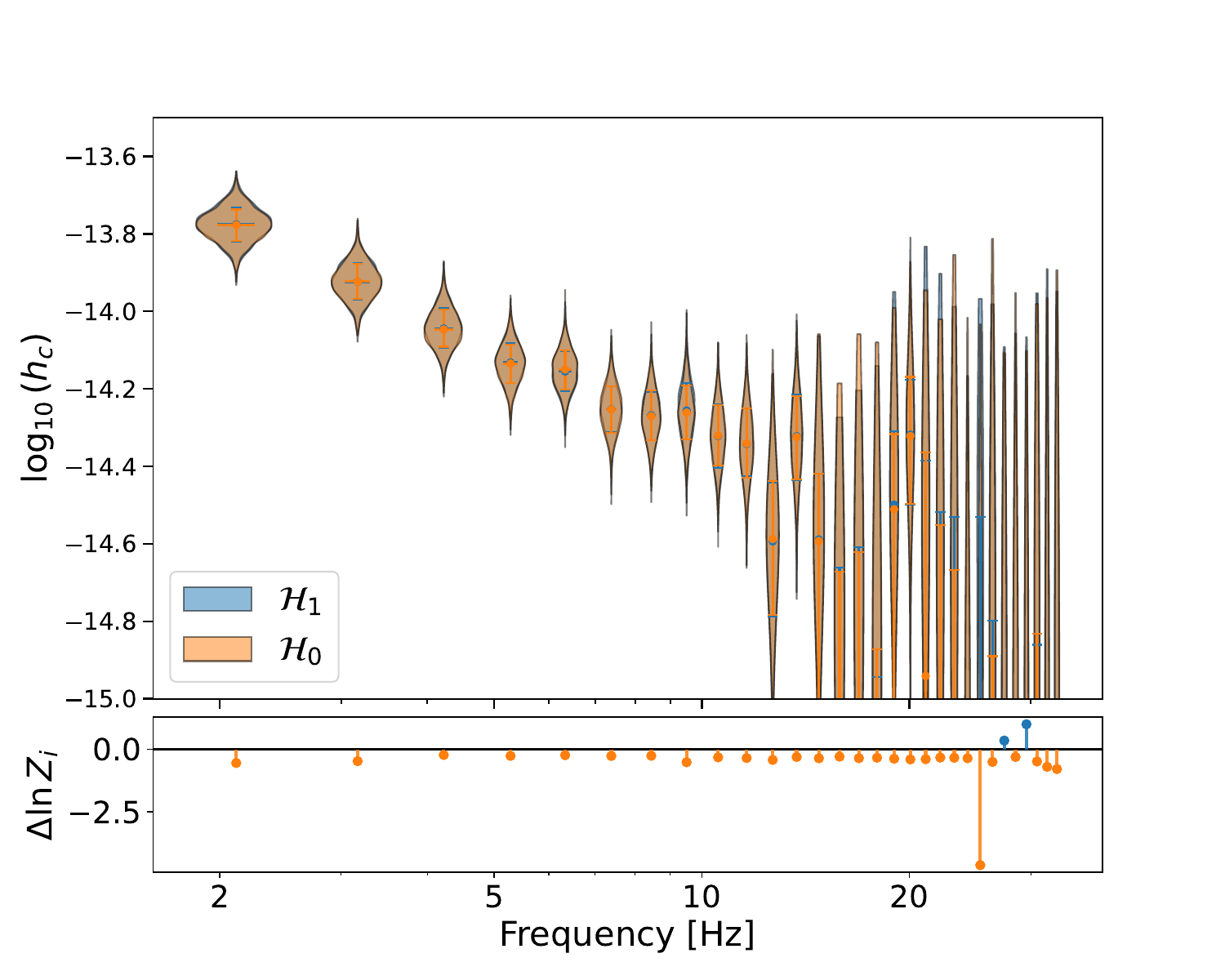}
    \caption{Same as Figs.~\ref{fig:spectrum}, but for $\gamma=2.5$.} 
    \label{fig:spectrum_2}
\end{figure}

We generate mock PTA data with the same PTA setup and population model except with a different power index $\gamma=2.5$. 
Posteriors of the free-spectrum reconstruction of the mock data are shown in Fig.~\ref{fig:spectrum_2}. Compared with the $\gamma=2.2$ case, the bin-wise evidence for an explicit bright-source component is much weaker. In most frequency bins, the data favor the GWB-only hypothesis $\mathcal{H}_0$, leading to total log Bayes factors  $\Delta_{(1)}\ln Z=-5.19$ from the 15 lowest frequency bins and $-13.79$ from all the 30 frequency bins (see Table~\ref{tab:bf_gamma25}).  This behavior is expected because the GWB with a steeper source-amplitude distribution is dominated by many fainter binaries, so the brightest source carries less power and the background is close to a Gaussian random signal with a smooth PSD.

\begin{figure}[h]
    \centering
   \includegraphics[width=8.cm]{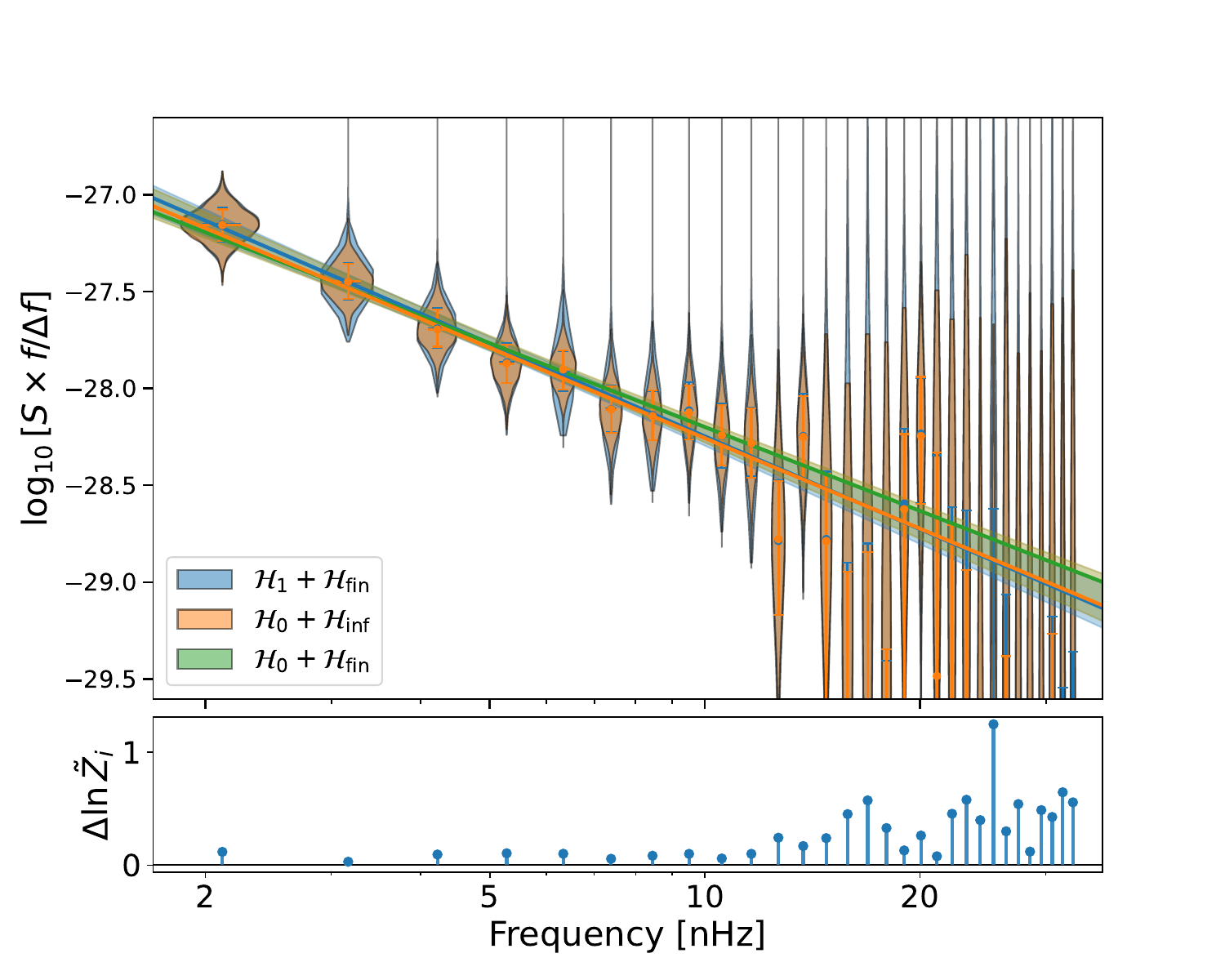}
    \caption{Same as Fig. \ref{fig:pop}, but for $\gamma=2.5$. }
    \label{fig:pop_2}
\end{figure}

The hierarchical population inference result is shown in Fig.~\ref{fig:pop_2}. The mixture hypothesis $\mathcal{H}_0+\mathcal{H}_{\rm fin}$ is almost indistinguishable from $\mathcal{H}_0+\mathcal{H}_{\rm inf}$ for the lowest 15 frequency bins, but is mildly favored when all 30 bins are included. After combining the two inference steps, the $\mathcal{H}_1+\mathcal{H}_{\rm fin}$ model is disfavored with log Bayes factors $\Delta\ln Z=-3.11$ and $-5.32$ for the 15- and 30-bin analyses, respectively. 

This example shows that even there is no evidence for bright sources in each frequency bin, the power spectrum fluctuations across different frequency bins also encode population 
information of a finite number of sources. This is one of the advantages of the hierarchical Bayesian inference consisting two steps: free spectrum reconstruction and Bayesian population inference. It clearly traces evidence for a finite population of GW sources, either from bright sources in each frequency bins and/or from power spectrum fluctuations across different frequency bins.

\begin{figure}[h]
    \centering
   \includegraphics[width=8.cm]{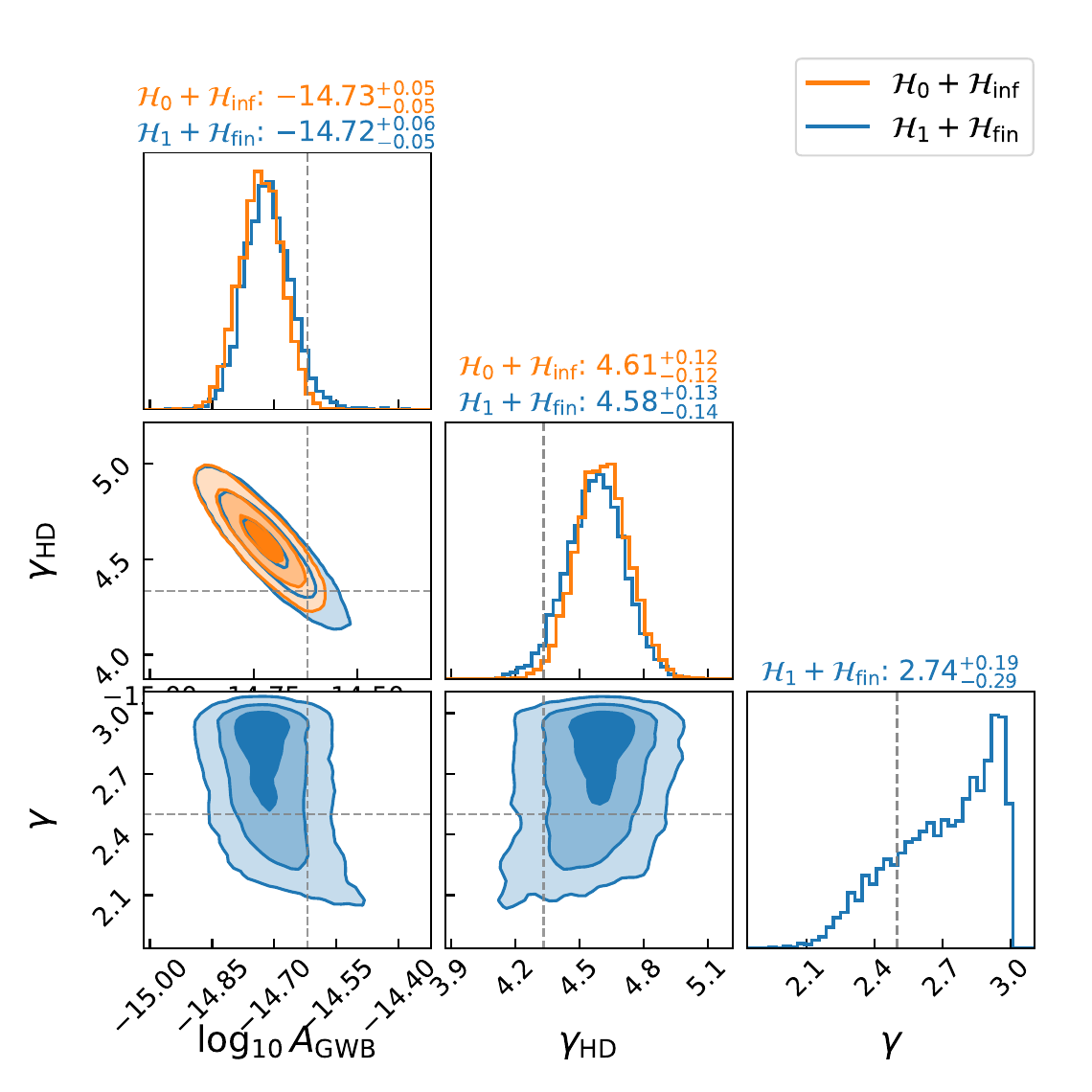}
    \caption{Same as Fig.~\ref{fig:corner}, but for $\gamma=2.5$. }
    \label{fig:corner_2}
\end{figure}

The posteriors of population model parameters are shown in the corner plot in Fig.~\ref{fig:corner_2}. They remain broadly consistent with the injected population parameters, 
where $A_{\rm GWB}$ and $\gamma_{\rm HD}$ are constrained by the overall trend of the PSD of the GWB, while $\gamma$ is constrained by fluctuations in the PSD. In the case of weak sources dominating the GWB, low $\gamma$ regime is excluded and high $\gamma$ regime is consistent with the low fluctuations in the PSD.



\end{document}